\documentclass[12pt]{article}

\usepackage{epsf}
\usepackage[dvips]{graphicx}

\begin{document}

\begin{minipage}{14cm}
\vskip 1cm
\hspace*{10.2cm} {\bf Preprint JINR}\\
\hspace*{10.5cm}{\bf E2-2005-76}\\
\hspace*{10.5cm}{\bf Dubna, 2005}\\
%\hspace*{10.1cm}{\bf hep-ph/......}\\
\end{minipage}
\vskip 1cm

\begin{center}
{\bfseries GENERALIZED Z-SCALING\\[2mm]
 FOR CHARGED HADRONS AND JETS}

\vskip 5mm
 I. Zborovsk\'{y}$^{{a},\natural}$
 %\footnote{E-mail:zborovsky@ujf.cas.cz}
 and  M. Tokarev$^{{b},\star}$
 %\footnote{E-mail:tokarev@sunhe.jinr.ru}

\vskip 0.2cm {\small
$^{(a)}${\it Nuclear Physics Institute,\\
Academy of Sciences of the Czech Republic, \\
\v {R}e\v {z}, Czech Republic} }

\vskip 5mm {\small
$^{(b)}${\it Veksler and Baldin Laboratory of High Energies,\\
Joint Institute for Nuclear Research,\\
141980, Dubna, Moscow region, Russia}}\\[3mm]
$^{\natural}${\it E-mail: zborov@ujf.cas.cz}\\
$^{\star}${\it E-mail: tokarev@sunhe.jinr.ru
}
\end{center}

\vskip 5mm

\begin{center}
\begin{minipage}{150mm}
\centerline{\bf Abstract}
 Generalization of $z$-scaling observed
 in the inclusive high-$p_T$
 charged hadron and jet  production is proposed.
The scaling function  $\psi(z)$ describing both charged hadrons
and jets produced in proton-(anti)proton collisions for various
multiplicity densities and collision energies is constructed.
Anomalous fractal dimensions and parameters characterizing
associated medium for both classes of events are established. The
basic features of the scaling established in minimum bias events
are shown to be preserved up to the highest multiplicity densities
measured in experiments UA1, E735, CDF and STAR. The obtained
results are of interest to use $z$-scaling as a tool for searching
for new physics phenomena of particle production in high
transverse momentum and high multiplicity region at the U70,
Tevatron, RHIC and LHC.
\end{minipage}
\end{center}

\vskip 20mm
\begin{center}
  Presented at the VIII International Workshop\\
  "Relativistic Nuclear Physics: From Mev to TeV",\\
VBLHE, JINR, May 23-28, 2004, Dubna, Russia\\[10mm]
Submitted to "Physics of Particles and Nuclei, Letters"
\end{center}

%\vskip 1cm
\newpage

{\section{Introduction}}

Collision of hadrons and nuclei at sufficiently high energies is an
ensemble of individual interactions of their constituents. The
constituents are partons in the parton model or quarks and gluons
in the theory of QCD. Production of particles with large
transverse momenta from such reactions has relevance to
constituent interactions at small scales. This regime is of
interest to search for new physical phenomena in elementary
processes such as quark compositeness \cite{Quarkcom}, extra dimensions \cite{Extradim},
black holes \cite{Blackhole}, fractal space-time \cite{Fracspace} etc.
Other aspects of high energy
interactions are connected with small momenta of secondary
particles and high multiplicities. This regime has relevance to
collective phenomena of particle production. Search for new
physics in both regions is one of the main goals of investigations
at Relativistic Heavy Ion Collider (RHIC) at BNL and Large Hadron
Collider (LHC) at CERN. Experimental data  of particle production
can give constraints for different theoretical models. Processes
with high transverse momenta of produced particles are most
suitable for precise test of perturbative QCD. The soft regime is
of interest for verification of non-perturbative QCD and
investigation of phase transitions in non-Abelian theories.

Many approaches to description of particle production are used to search for
regularities reflecting general principles at high energies
\cite{Feynman}-\cite{Brodsky}.
 One
of the most basic principles is the self-similarity of hadron
production valid both in soft and hard physics. Other general
principles are locality and fractality which can be applied to
hard processes at small scales. The locality of hadronic
interactions follows from numerous experimental and theoretical
studies. These investigations shown that interactions of hadrons
and nuclei can be described in terms of the interactions of their
constituents. Fractality in hard processes is specific feature
connected with sub-structure of the constituents. This includes the
self-similarity  over wide scale range. Fractality of soft
processes concerning the multi-particle production was
investigated by many authors.

Fractality in inclusive reactions with high-$p_T$ particles
was considered for the first time in the framework of $z$-scaling\cite{Z}.
The approach is based on
principles of fractality, locality and self-similarity. It takes
into account fractal structure of the colliding objects, interaction of
their constituents and particle formation. The observed regularity
is expressed in terms of the scaling function $\psi(z)$ and the
scaling variable $z$ constructed via the experimentally
measured inclusive cross section $Ed^3\sigma/dp^3$ and the
multiplicity density $dN/d\eta$. Data $z$-presentation reveals
properties of energy and angular independence and power law at
high $z$. The function $\psi(z)$ is interpreted as probability
density to produce a particle with formation length $z$.

In the present paper we generalize the concept of $z$-scaling for
various multiplicities of produced particles. We show that
generalized scaling represents regularity in both soft and hard
regime in proton-(anti)proton collisions over a wide range of multiplicities.
Connection of the $z$-scaling with entropy and heat capacity of the colliding system is
discussed. The construction is applied to experimental data on
inclusive cross sections of charged hadrons and jets.
The same scaling function $\psi(z)$ was established to describe
production of both classes of events.
The analysis of data shows simultaneously that anomalous fractal
dimensions and "heat capacity"  for charged
hadrons and jets are different. This indicates that production mechanism
is  influenced differently by the system at hadron and parton level.
The self-similarity of the production processes governed
by the same law is common to both of them.
The obtained results give additional confirmation of self-similarity and fractality in
hadronic processes at high energies. We suggest to use $z$-scaling
as a tool for searching for new physics phenomena of particle
production in high transverse momentum and high multiplicity
region at the U70, Tevatron, RHIC and LHC.

\vskip 0.5cm
{\section{$Z$-Scaling}}

We would like to emphasize two main points of our approach \cite{Z}. The
first one is based on self-similarity of the mechanism
underlying particle production on the level of the elementary
constituent interactions.
The interactions are governed by local energy-momentum conservation law.
The second one is fractal character
of the parton content of the composite structures involved.

\vskip 0.5cm
{\subsection{Locality, self-similarity and fractality}}

The idea of $z$-scaling is based on the assumption \cite{Stavinsky}
that gross features of inclusive particle distribution of the
reaction
\begin{equation}
M_{1}+M_{2} \rightarrow m_1 + X
\label{eq:r1}
\end{equation}
can be described at high energies in terms of the corresponding
kinematic characteristics of constituent sub-processes.
The sub-process is considered as binary collision
\begin{equation}
(x_{1}M_{1}) + (x_{2}M_{2}) \rightarrow m_1/y +
(x_{1}M_{1}+x_{2}M_{2} + m_2/y)
\label{eq:r2}
\end{equation}
of constituents which
carry the fractions $x_1$ and $x_2$ of the incoming 4-momenta $P_1$ and $P_2$
of the objects with the masses $M_1$ and $M_2$.
The inclusive particle with the mass $m_1$ and the 4-momentum $p$
carries out the fraction $y$ of the 4-momentum of the outgoing constituent.
The constituent interaction satisfies the energy-momentum conservation written in the form
\begin{equation}
(x_1P_1+x_2P_2-p/y)^2 =(x_1M_1+x_2M_2+m_2/y)^2.
\label{eq:r3}
\end{equation}
The parameter $m_2$ is
introduced to satisfy the internal conservation laws (for baryon
number, isospin, strangeness, and so on).
The equation is the expression of locality of  hadron interaction
at constituent level.

The constituent interactions are assumed to be similar.
This property is connected with  dropping of certain dimensional
quantities out of description of physical phenomena.
The self-similar solutions are constructed in terms of self-similarity parameters.
We search for the solution
\begin{equation}
\psi(z) ={1\over{N\sigma}}{d\sigma\over{dz}}
\label{eq:r4}
\end{equation}
depending on single self-similarity parameter $z$. Here $\sigma$ is the inelastic cross
section of the reaction (\ref{eq:r1}) and $N$ is particle multiplicity.
The  parameter $z$ is specific dimensionless combination
of  quantities  which  characterize particle production in high energy
inclusive  reactions. It depends on  momenta and masses of  the colliding
and  inclusive particles,  multiplicity density and structural characteristics of
the interacting objects.
We define the self-similarity parameter $z$ in the form
\begin{equation}
z =z_0 \Omega^{-1}
\label{eq:r5}
\end{equation}
where
\begin{equation}
\Omega(x_1,x_2,y)=(1-x_1)^{\delta_1}(1-x_2)^{\delta_2}(1-y)^{\epsilon}.
\label{eq:r6}
\end{equation}
The  parameter $z$ has character of a fractal measure.
For a given production process, its finite part $z_0$ is proportional to the
kinetic transverse energy  released in the underlying  collision of
constituents. The divergent part $\Omega^{-1}$ describes the resolution
at which the collision of the constituents can be singled out of this process.
The $\Omega(x_1,x_2,y)$ is relative number of parton configurations containing
constituents which carry the fractions $x_1$ and  $x_2$ of the incoming momenta
$P_1$ and $P_2$  and the outgoing constituent which fraction $y$ is carried out
by the inclusive particle with the momentum $p$.
The $\delta_1$, $\delta_2$ and $\epsilon$ are anomalous fractal dimensions
of the incoming and outgoing objects, respectively.
Common property of fractal measures is their divergence with the increasing resolution
\begin{equation}
z(\Omega) \rightarrow \infty \ \ \ \ \ \ \ \ if  \ \ \ \ \  \Omega^{-1} \rightarrow \infty.
\label{eq:r7}
\end{equation}
For the infinite resolution, the momentum fractions become unity, $x_1=x_2=y=1$ and $\Omega=0$.
The kinematical limit corresponds to the fractal limit $z=\infty$.

\vskip 0.5cm
{\subsection{Principle of minimal resolution and momentum fractions}}

Let us formulate principle of minimal resolution $\Omega^{-1}$ of the fractal
measure $z$ with respect to all
binary collisions of constituents from which the inclusive particle with
 the momentum $p$ can be produced.
This singles out the underlying sub-process of the constituents with
the resolution beyond which there is no sense
to consider their sub-structure.
In such a way, the momentum fractions $x_{1}$, $x_{2}$ and $y$ are
determined to minimize  $\Omega^{-1}(x_{1},x_{2},y)$
taking into account the energy-momentum conservation
in the binary collision (\ref{eq:r2}).
This is equivalent to the solution of the system of equations
\begin{equation}
\frac{\partial\Omega(x_1,x_2,y)}{\partial x_1} = 0, \ \ \ \ \ \
\frac{\partial\Omega(x_1,x_2,y)}{\partial x_2} = 0, \ \ \ \ \ \
\frac{\partial\Omega(x_1,x_2,y)}{\partial y} = 0
\label{eq:r8}
\end{equation}
with the condition (\ref{eq:r3}).
The principle of minimal resolution leads to the decomposition of the momentum fractions
\begin{equation}
x_1=\lambda_1/y+\chi_1(\alpha,y)/y \ \ \ \ \ \ \
x_2=\lambda_2/y+\chi_2(\alpha,y)/y
\label{eq:r10}
\end{equation}
at fixed $y$. Using the decomposition, the expression (\ref{eq:r2}) can be rewritten to the symbolic form
\begin{equation}
x_1+x_2\rightarrow (\lambda_1+\lambda_2)/y + (\chi_1+\chi_2)/y.
\label{eq:r11}
\end{equation}
This relation means that $\lambda$-parts of the interacting partons contribute to the production
of the inclusive particle, while the $\chi$-parts are responsible for the creation of its recoil.
Here we use the following notations
\begin{equation}
\chi_1=\sqrt{\mu_1^2+\omega_1^2}-\omega_1 \ \ \ \
\chi_2=\sqrt{\mu_2^2+\omega_2^2}+\omega_2
\label{eq:r12}
\end{equation}
where
\begin{equation}
\mu_1^2=(\lambda_1\lambda_2+\lambda_0)\alpha\frac{y-\lambda_1}{y-\lambda_2} \ \ \ \
\mu_2^2=(\lambda_1\lambda_2+\lambda_0)\alpha^{-1}\frac{y-\lambda_2}{y-\lambda_1}.
\label{eq:r13}
\end{equation}
The parameter $\alpha=\delta_2/\delta_1$ is the ratio of the anomalous fractal dimensions of the colliding
objects and
\begin{equation}
\lambda_1=\frac{(P_2p)+M_2m_2}{(P_1P_2)-M_1M_2}, \ \ \ \
\lambda_2=\frac{(P_1p)+M_1m_2}{(P_1P_2)-M_1M_2}, \ \ \ \
\lambda_0=\frac{0.5(m_2^2-m_1^2)}{(P_1P_2)-M_1M_2}.
\label{eq:r14}
\end{equation}
The $\omega_i=\mu_iU (i=1,2)$ are expressed through the quantity
\begin{equation}
U = \frac{\alpha-1}{2\sqrt{\alpha}}\xi
\label{eq:r15}
\end{equation}
with the kinematical factor
\begin{equation}
\xi=\sqrt{\frac{\lambda_1\lambda_2+\lambda_0}{(y-\lambda_1)(y-\lambda_2)}},
\label{eq:r16}
\end{equation}
$(0\le\xi\le 1)$, characterizing scale of the constituent interaction.
Solution of the system (\ref{eq:r8}) with the condition (\ref{eq:r3}) can be obtained by iterations
of expressions (\ref{eq:r10}) with respect to $y$.

\vskip 0.5cm
{\subsection{Scaling variable $z$}}

Search for an adequate, physically meaningful but still sufficiently simple form of the
self-similarity parameter $z$ plays a crucial role in our approach.
We define the scaling variable $z$ in the form
\begin{equation}
z = \frac{s^{1/2}_{\bot}}{(dN/d\eta|_0)^c \cdot m_0}\cdot\Omega^{-1}.
\label{eq:r17}
\end{equation}
Here $m_0$ is the  nucleon mass.
Minimal transverse kinetic energy of the constituent sub-process is determined
by the formula
\begin{equation}
s^{1/2}_{\bot}= s^{1/2}_{\lambda} +  s^{1/2}_{\chi} - m_1-(M_1x_1y+M_2x_2y+m_2)
\label{eq:r18}.
\end{equation}
This energy consists of two parts
\begin{equation}
s^{1/2}_{\lambda}=\sqrt{(\lambda_1P_1+\lambda_2P_2)^2},     \ \ \ \ \
s^{1/2}_{\chi}=\sqrt{(\chi_1P_1+\chi_2P_2)^2}
\label{eq:r19}
\end{equation}
which represent the energy for the creation of the inclusive particle and its recoil, respectively.
The boundaries of the range of $z$ are 0 and $\infty$. They are accessible
at any collision energy and do not depend on the kinematical factor $\xi$.

The $dN/d\eta|_0$ is multiplicity density at (pseudo)rapidity $\eta=0$.
It depends on state of the produced medium in the colliding system.
The parameter $c$ characterizes properties of this medium.
The quantity
\begin{equation}
{\it W}=(dN/d\eta|_0)^c\cdot\Omega
\label{eq:r20}
\end{equation}
is proportional to all parton and hadron configurations of the colliding system
resulting to production of the inclusive particle with the momentum $p$.
The scaling variable (\ref{eq:r17}) has physical meaning of the ratio
\begin{equation}
z = \frac{s^{1/2}_{\bot}}{{\it W }\cdot m_0}
\label{eq:r20a}
\end{equation}
of the minimal transverse kinetic energy of the constituent sub-process
and the number of the configurations $W$.

\vskip 0.5cm
{\subsection{Scaling variable $z$ and entropy ${\it S}$}}

Let us introduce correspondence between the scaling variable and entropy.
According to statistical physics, entropy of a system is given by
number of all statistical states $W$ of the system as follows
\begin{equation}
{\it S} = \ln {\it W}.
\label{eq:r20b}
\end{equation}
In thermodynamics, entropy for ideal gas is determined by the formula
\begin{equation}
{\it S} = c_V\ln {T}+ R\ln {V} + const.
\label{eq:r20c}
\end{equation}
The $c_V$ is heat capacity and $R$ is universal constant. The temperature $T$
and the volume $V$ characterize state of the system. Using
(\ref{eq:r20}) and (\ref{eq:r20b}), we can write
\begin{equation}
{\it S} = c\ln {\left[dN/d\eta|_0\right]}+
\ln{[(1\!-\!x_1)^{\delta_1}(1\!-\!x_2)^{\delta_2}(1\!-\!y)^{\epsilon}]} + const.
\label{eq:r20d}
\end{equation}
Exploiting analogy between Eqs. (\ref{eq:r20c}) and (\ref{eq:r20d}), we interpret
the parameter $c$ as "heat capacity" of the medium. The multiplicity density $dN/d\eta|_0$
 has physical meaning of "temperature" of the colliding system.
The second term in Eq. (\ref{eq:r20d}) depends on the volume in space of the
momentum fractions $\{x_1,x_2,y\}$ containing such constituent configurations which
can contribute to production of the inclusive particle with the momentum $p$.
On basis of this analogy we can say that entropy of the colliding system
increases with the multiplicity density and decreases with
increasing resolution $\Omega^{-1}$.

\vskip 0.5cm
{\subsection{Scaling function $\psi(z)$}}

The scaling function $\psi(z)$  is expressed in terms of the experimentally
measured inclusive invariant cross section $Ed^3\sigma/dp^3$, multiplicity density
$dN/d\eta$ and the total inelastic cross section $\sigma_{in}$.
Exploiting the definition (\ref{eq:r4}) one can obtain the expression
\begin{equation}
\psi(z) = -{ { \pi s} \over { (dN/d\eta) \sigma_{in}} } J^{-1} E {
{d^3\sigma} \over {dp^3}  }
\label{eq:r21}
\end{equation}
Here, $s$ is the center-of-mass collision energy squared and
\begin{equation}
J = \frac{\partial y }{\partial\lambda_1}\frac{\partial z }{\partial\lambda_2}-
\frac{\partial y }{\partial\lambda_2}\frac{\partial z }{\partial\lambda_1}
\label{eq:r22}
\end{equation}
is the corresponding Jacobian. The rapidity is given by
\begin{equation}
y=\frac{1}{2}\ln \frac{\lambda_2}{\lambda_1}.
\label{eq:r23}
\end{equation}
The function $\psi(z)$ is normalized as follows
\begin{equation}
\int_{0}^{\infty} \psi(z) dz = 1.
\label{eq:r24}
\end{equation}
The relation allows us to interpret the $\psi(z)$ as a
probability density to produce inclusive particle with the corresponding
value of the variable $z$. The variable $z$ was interpreted  as a particle formation length.

\vskip 0.5cm
{\section{Multiplicity dependence of $\psi(z)$ for charged hadrons}}

We analyze experimental data \cite{UA1,E735,CDF,STAR}
on charge hadron production in $p\bar{p}$  and $pp$ collisions
at different multiplicities and energies over a wide $p_T$ range.

The transverse momentum distributions of charged hadrons produced in  $p\bar{p}$ collisions
were measured by the UA1 Collaboration at the energy $\sqrt{s}=540~$GeV for
$dN/d\eta=0.8-14.9$ at the  SpS \cite{UA1}.
The measurements covered 5 units of  pseudorapidity in the central region of the
collisions, $|\eta|<2.5$. Multiplicity dependence of the transverse momentum distributions
were measured up to 6~GeV/c. Figure 1(a) demonstrates strong sensitivity of the spectra
to the multiplicity density at high $p_T$.
The same data are presented in Figure 1(b) in the scaling form. The scaling function $\psi(z)$
changes over 6 orders of magnitude in the range of $z=0.1-15$. The independence of $\psi$ on
multiplicity density $dN/d\eta$ is observed. The result gives strong restriction on
the parameter $c$. It was found to be $c=0.25$.

The E735 Collaboration measured the multiplicity dependence of charged hadron spectra
 in  proton-antiproton collisions at the energy $\sqrt{s}=1800~$GeV
for $dN/d\eta=2.3-26.2$ at the Tevatron \cite{E735}.
 This includes measurements at highest multiplicity
density and energy. The pseudorapidity range was $|\eta|<3.25$. Data cover the transverse
momentum range $p_T = 0.15-3~$GeV/c. Figure 2(a) confirms strong dependence of the spectra
on the multiplicity observed at the SpS. Figure 2(b) shows the same data in the $z$-presentation.
The independence of the scaling function $\psi(z)$ was found with the same value of $c=0.25$.
This includes region of small $z$ up to $0.03$.

New data on the multiplicity dependence on charged hadron spectra were obtained by the CDF
Collaboration at the Tevatron \cite{CDF}. The data were taken at collision energies
$\sqrt{s}=630$ and 1800~GeV. The particles were registered in the pseudorapidity
range $|\eta|<1$. The spectra were measured up to 10~GeV/c. Figures 3(a) and 3(b) show spectra
for multiplicity densities 2.5, 5.0 and 7.5. The multiplicity dependence of the spectra
increases with $p_T$. It is additional confirmation of the same effect observed
by the UA1 and E735 Collaborations.
Both CDF data sets are plotted in $z$-presentation in Figure 3(c). The scaling function reveals
the energy and multiplicity independence at the same value of $c=0.25$.
We note that the scaling for charged particle production in proton-antiproton collisions
for different multiplicities and energies is consistent with the values of the anomalous
fractal dimensions $\delta_1=1$, $\delta_2=1$ and $\epsilon=1$.

The STAR Collaboration obtained the new data \cite{STAR} on the inclusive spectrum
of charged hadrons produced in $pp$ collisions in the central rapidity range $|\eta|<0.5$
at the energy $\sqrt s = 200$~GeV.  The transverse momentum spectra were measured up to 9.5~GeV/c.
Figure 4(a) demonstrates the strong dependence of the spectra on multiplicity density
over the range $dN/d\eta=2.5, 6.0$ and 8.0.
The STAR data for proton-proton collisions at the RHIC confirm the multiplicity independence
of the scaling function established for proton-antiproton collisions at higher energies.
It was also found that the value of the parameter $c$ is equal to 0.25.

Thus we can conclude that the available experimental data on the multiplicity dependence
of charged particle spectra at various energies in $p\bar{p}$ and $pp$ collisions confirm
generalized $z$-scaling for the same value of the parameter $c$.

\vskip 0.5cm
{\section{Scaling function for charged hadrons and jets}}

Production of charged hadrons with large transverse momenta
has deep correspondence with the jet production. Jets are collimated groups of
hadrons including a leading particle with high enough $p_T$ and
associated hadrons. Jets are considered as direct manifestation of
hard parton scattering.

Charged hadron spectra were measured over a wide range of collision energies
$\sqrt{s}=200-1800$~GeV in proton-antiproton collisions. The transverse momentum
range covered 0.2-26~GeV/c. The experimental data on inclusive cross sections obtained
at the SpS and Tevatron are shown in Figure 5(a). The cross sections of jets are much smaller.
The D0 Collaboration data on jet production at $\sqrt{s}=630$ and 1800~GeV are plotted
in the same Figure. Note the strong dependence of cross sections on the collision energy
which increases with transverse momentum. The large values of transverse momenta up to
$p_T=450$~GeV/c correspond to interactions at very small scales.

It is possible to obtain the $z$-presentation of the charged hadron and jet spectra
with the same scaling function $\psi(z)$ shown in Figure  5(b). The values of $\psi$
vary over 15 orders of magnitude corresponding to the range $z=0.2-1000$.
We observe well matching of the scaling function for charged hadrons and jets in
large overlapping region.
The shape of the scaling curve is different in low and high $z$ region. The first one
is characteristic for soft regime of particle production.
The power behavior, $\psi(z)\sim z^{-\beta}$, seen for $z>4$ is typical for hard processes.
The common description was obtained for different values of fractal anomalous
dimension $\epsilon$ and parameter $c$ for both classes of events.
We found $\epsilon=1$, $c=0.25$ for charged hadrons and $\epsilon=0$, $c=1$
for jets. The fractal dimensions $\delta_1$ and $\delta_2$ are equal to 1 for both cases.

Let us discuss the obtained values of the parameters $\epsilon$ and $c$.
The zero value of $\epsilon$ for jet production means that all momentum of the
outgoing constituent is carried out by jet $(y=1)$.
In the case of inclusive hadron production, only part of the constituent momentum
$(y<1)$ is transferred to the detected particle.
This is reflected by the non-zero anomalous dimension $\epsilon=1$.
Results of our analysis show jump of the "heat capacity" $c$ from 0.25 for charged hadrons
to 1 for jets. This reflects transition from hadron to quark and gluon degrees
of freedom. Strong correlation between the "heat capacity" and the anomalous fractal
dimension $\epsilon$ is observed.
Final states, $hadron$ and $parton \equiv jet$,
characterized by these parameters, are different for the two classes of events.

\vskip 0.5cm
{\section{Conclusions}}

Dependence of charged hadron spectra on multiplicity density
measured by the UA1, E735 and CDF Collaborations in $\bar pp$
collisions  was studied in the framework of the $z$-scaling
concept. $Z$-presentation of the data reveal basic features
established in minimum bias events. We showed that beside energy
independence, the scaling function $\psi(z)$ indicates
multiplicity independence in the whole measured $dN/d\eta$ range.
New experimental data on multiplicity dependencies of charged
hadron spectra  in $pp$ collisions obtained by the STAR
Collaboration at RHIC additionally confirm the observed results.

Generalized scaling for the  inclusive charged hadron and jet
production was suggested. It was shown that the same scaling
function $\psi(z)$ expressed via the invariant inclusive cross
section $Ed^3\sigma/dp^3$  and the charged hadron multiplicity
density $dN/d\eta|_0$  describes both classes of events.
%The scaling variable $z$ is ratio of the minimal transverse kinetic
%energy of the underlying sub-process  and relative number of all
%configurations of the colliding system in which the inclusive
%particle/jet can be created.
The variable $z$ has property of a fractal measure  connected
with parton content of the composite structures involved.
The constituent sub-process underlying
production of the inclusive particle/jet is determined by minimal
resolution of the fractal measure $z$ with respect to all
sub-processes in which the inclusive particle/jet can be produced.

The anomalous fractal dimensions and the parameter $c$ for charged
hadrons and jets were found to be $\delta_1=\delta_2=1,
\epsilon=1, c=0.25$ and $\delta_1=\delta_2=1, \epsilon=0, c=1$,
respectively. Connection between the scaling variable and entropy
was used to interpret the parameter $c$ as "heat capacity" of the
produced medium and the multiplicity density  in the central
region $dN/d\eta|_0$ as "temperature" of the colliding system.
Analysis of experimental data on charged particle production in
$p\bar{p}$ and $pp$ collisions indicates constancy of the "heat
capacity" $c$ for various energies and multiplicities. Jump of the
"heat capacity" from 0.25 for charged hadrons to 1 for jets was
argued as indication of phase transition from hadron to
quark-gluon degrees of freedom.

The obtained results are of interest for searching and study of
new physics phenomena in particle production over a wide range of
collision energies and multiplicity densities at the U70,
Tevatron, RHIC and LHC.

\vskip 5mm
{\large \bf Acknowledgments.} The authors would like to
thank  Yu.Panebratsev for his support of this work. The
investigations have been partially supported by the IRP
AVOZ10480505 and by the Grant Agency of the Czech Republic under
the contract No. 202/04/0793.

% *************    1(a,b) 2(a,b)  *************************
\newpage
\begin{minipage}{4cm}

\end{minipage}

\vskip 4cm
\begin{center}
\hspace*{-2.5cm}
\parbox{5cm}{\epsfxsize=5.cm\epsfysize=5.cm\epsfbox[95 95 400 400]
{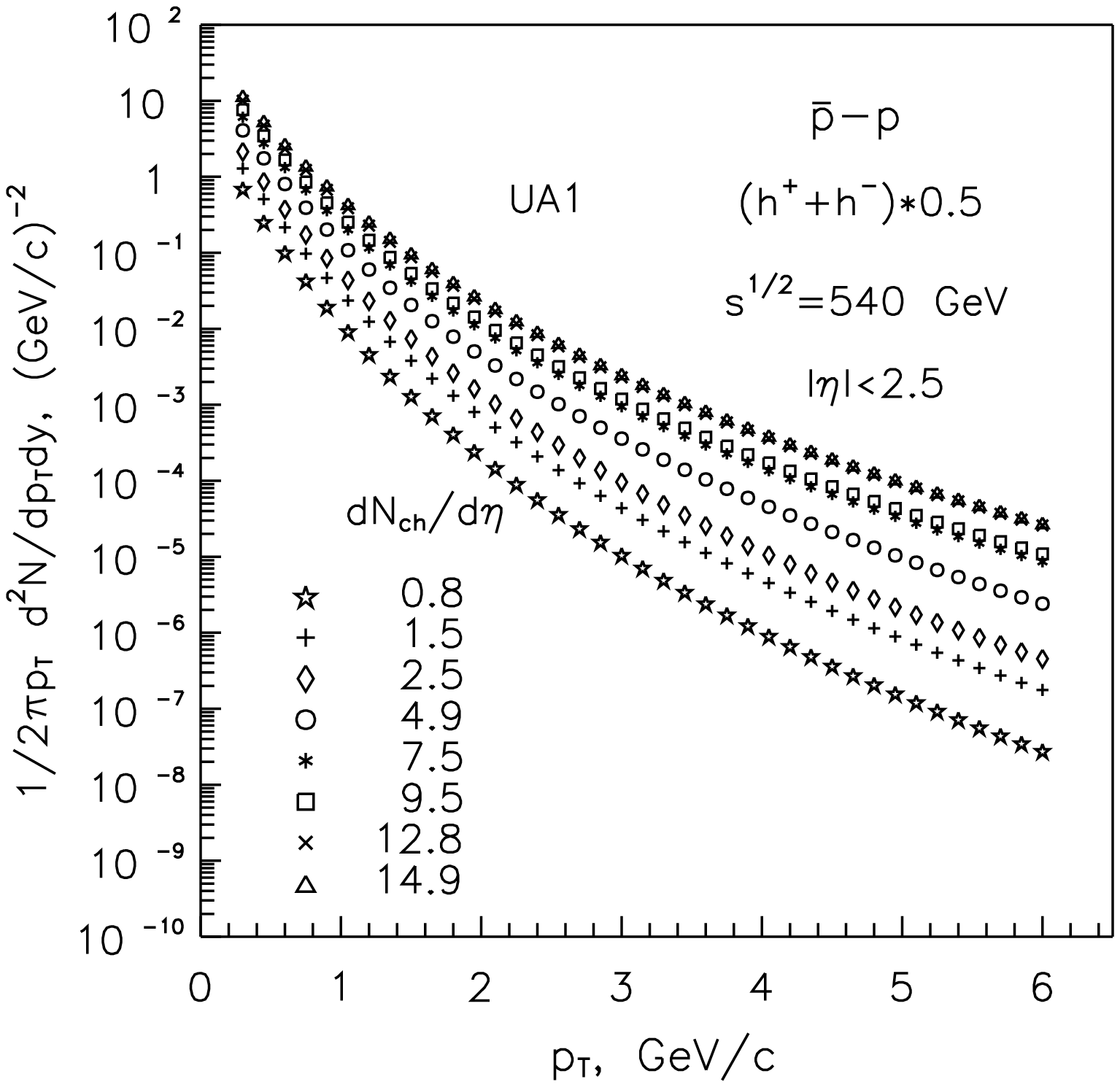}{}}
\hspace*{3cm}
\parbox{5cm}{\epsfxsize=5.cm\epsfysize=5.cm\epsfbox[95 95 400 400]
{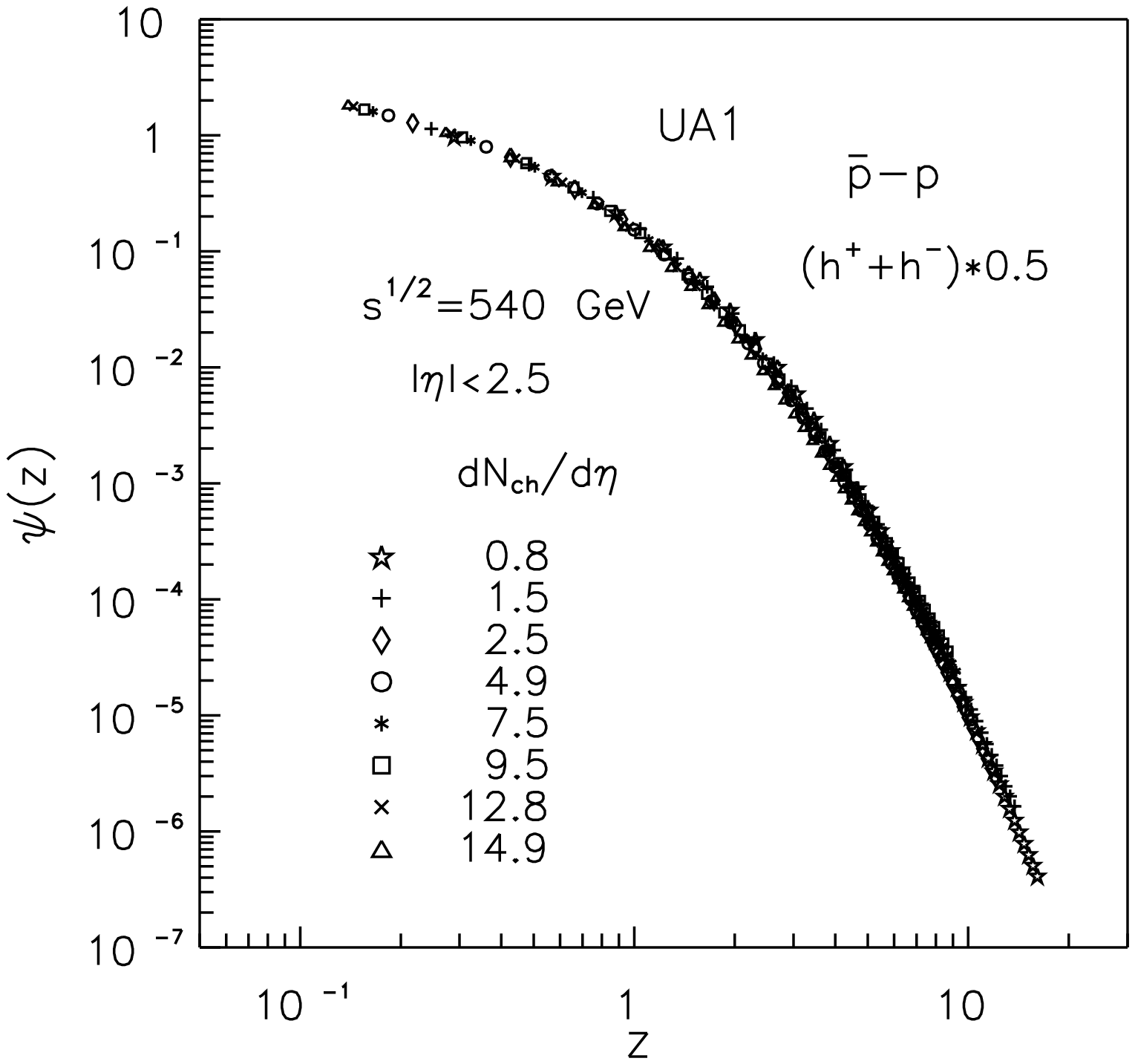}{}}
\vskip -1.cm
\hspace*{0.cm} a) \hspace*{8.cm} b)\\[0.5cm]
\end{center}

{\bf Figure 1.}
(a) Multiplicity dependence of charged hadron spectra
  in $\bar pp$ collisions  at $\sqrt s=540$~GeV.
 Experimental data are obtained by the UA1 Collaboration \cite{UA1}.
(b) The corresponding scaling function $\psi(z)$.

\vskip 5cm

\begin{center}
\hspace*{-2.5cm}
\parbox{5cm}{\epsfxsize=5.cm\epsfysize=5.cm\epsfbox[95 95 400 400]
{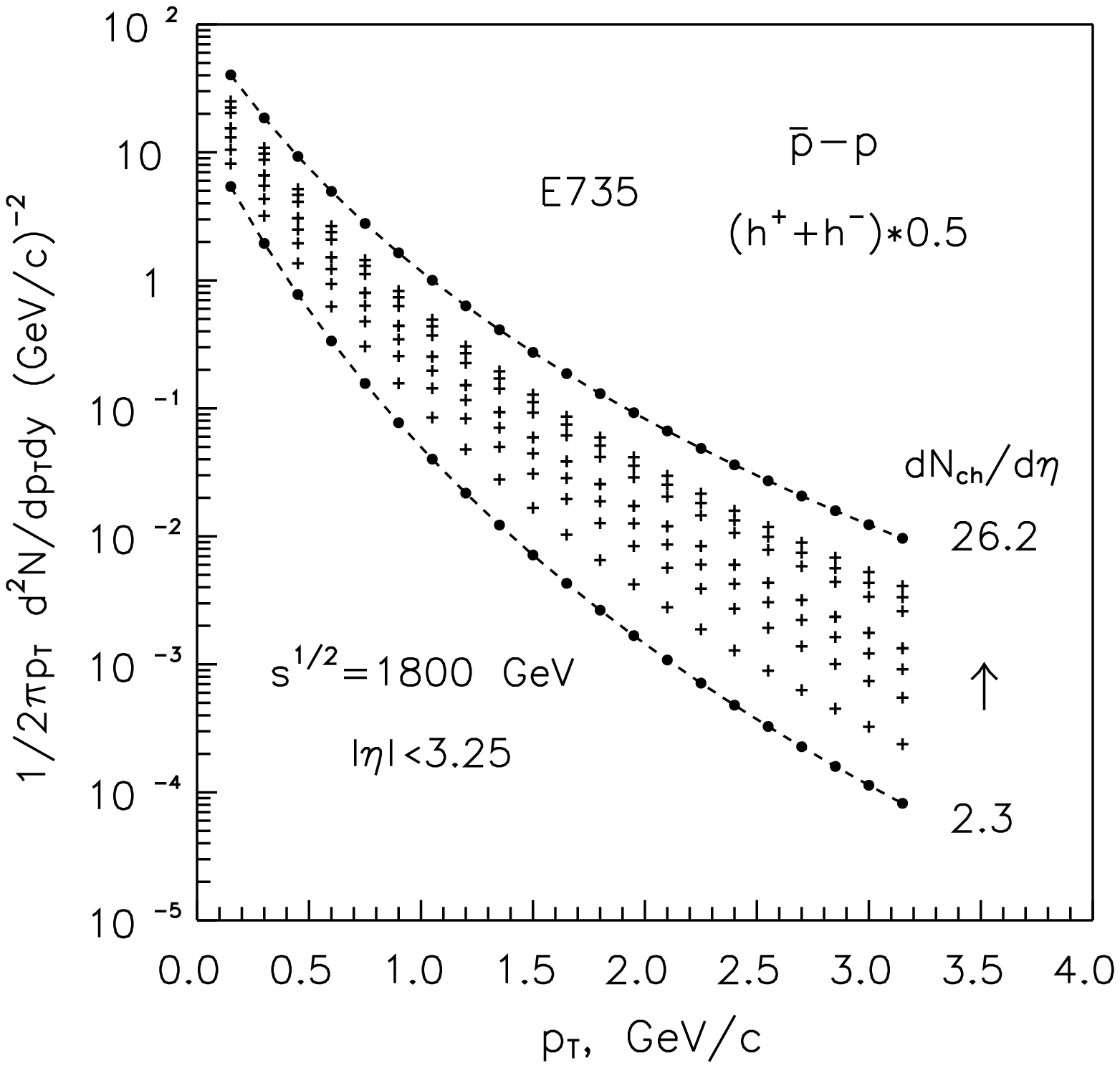}{}}
\hspace*{3cm}
\parbox{5cm}{\epsfxsize=5.cm\epsfysize=5.cm\epsfbox[95 95 400 400]
{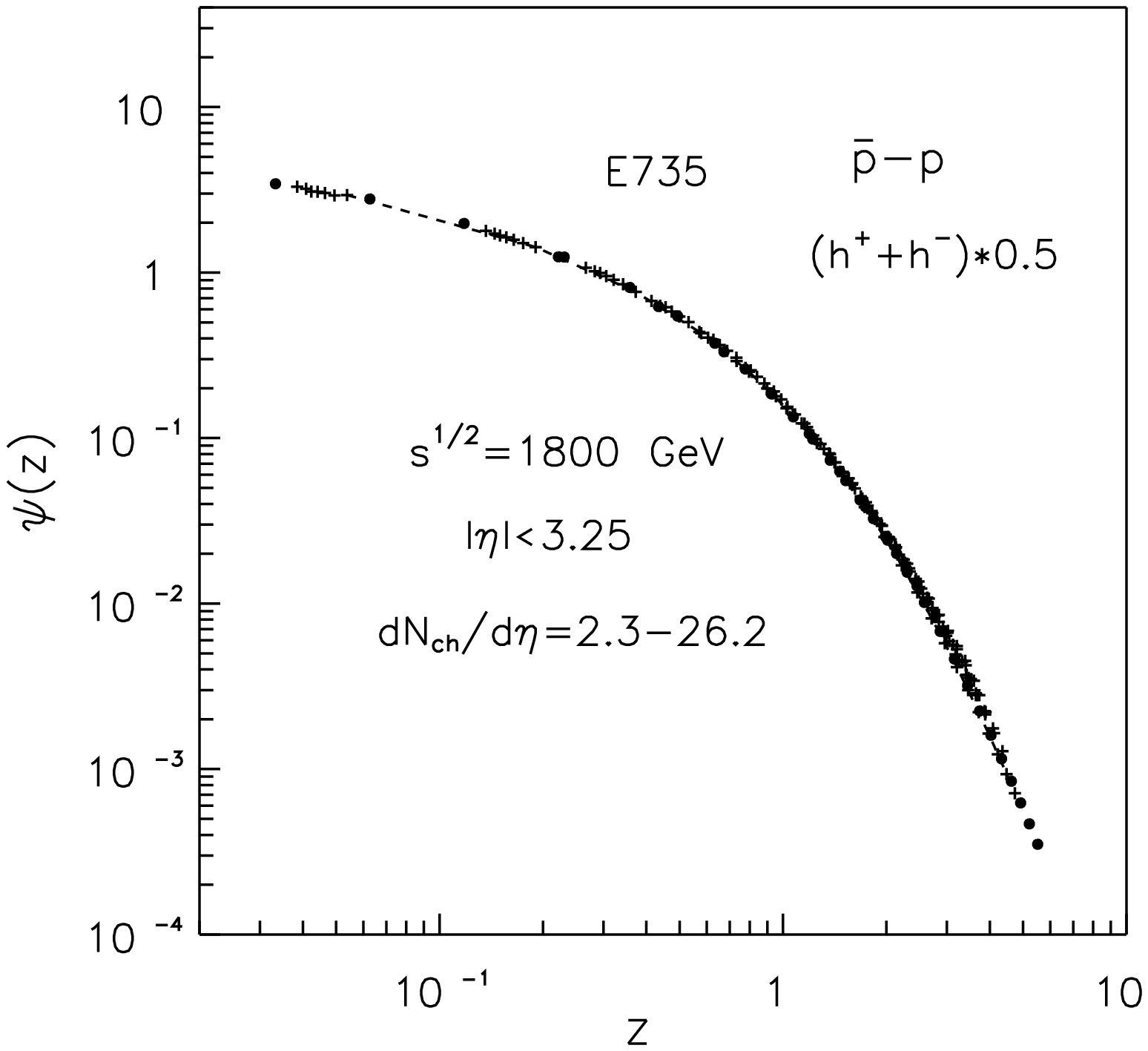}{}}
\vskip -1.cm
\hspace*{0.cm} a) \hspace*{8.cm} b)\\[0.5cm]
\end{center}

{\bf Figure 2.}
a) Multiplicity dependence of charged hadron spectra
  in $\bar pp$ collisions  at $\sqrt s=1800$~GeV.
 Experimental data are obtained by the E735 Collaboration \cite{E735}.
(b) The corresponding scaling function $\psi(z)$.

% *************    3(a,b,c)  *************************
\newpage
\begin{minipage}{4cm}

\end{minipage}

\vskip 4cm
\begin{center}
\hspace*{-2.5cm}
\parbox{5cm}{\epsfxsize=5.cm\epsfysize=5.cm\epsfbox[95 95 400 400]
{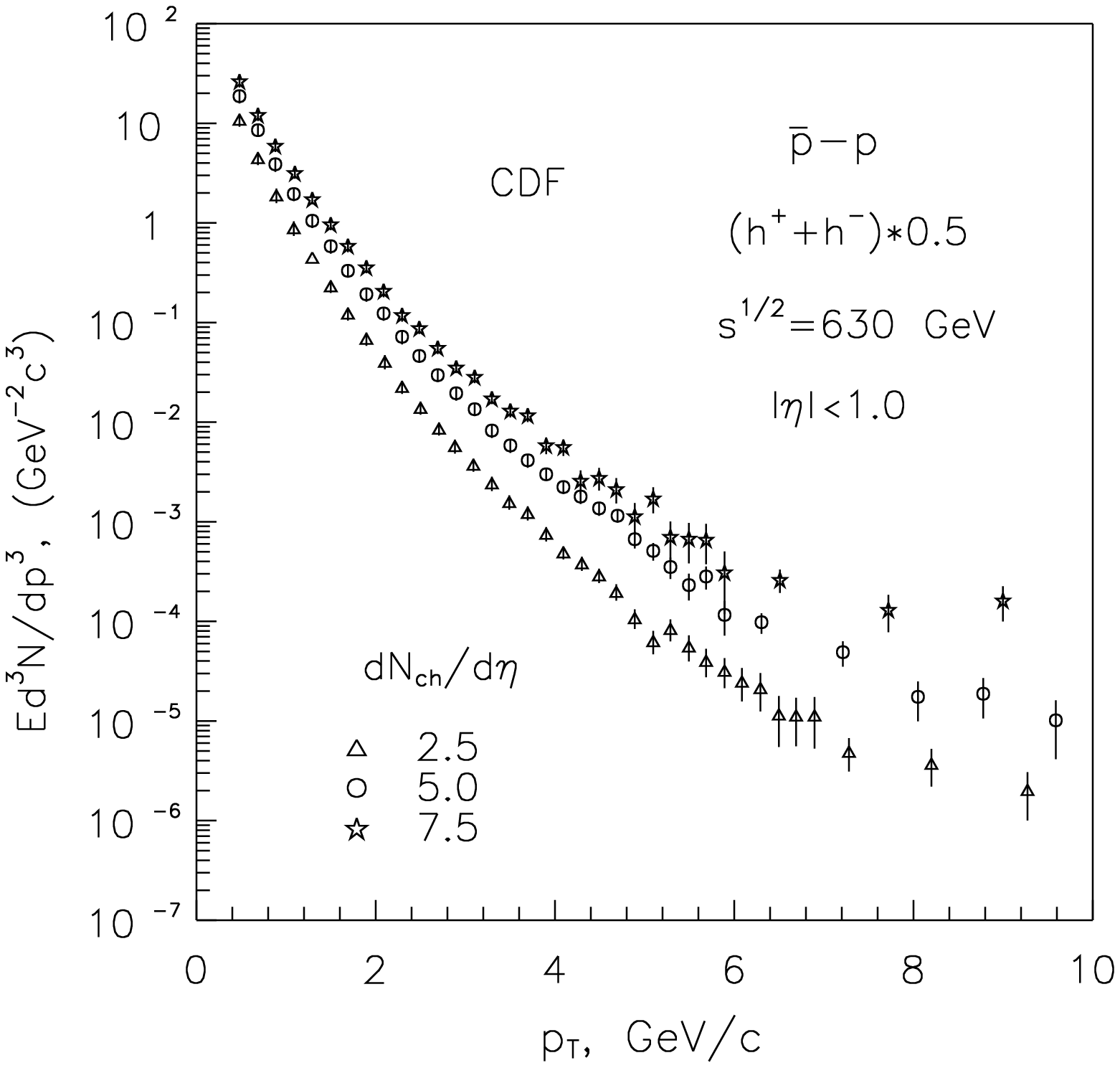}{}}
\hspace*{3cm}
\parbox{5cm}{\epsfxsize=5.cm\epsfysize=5.cm\epsfbox[95 95 400 400]
{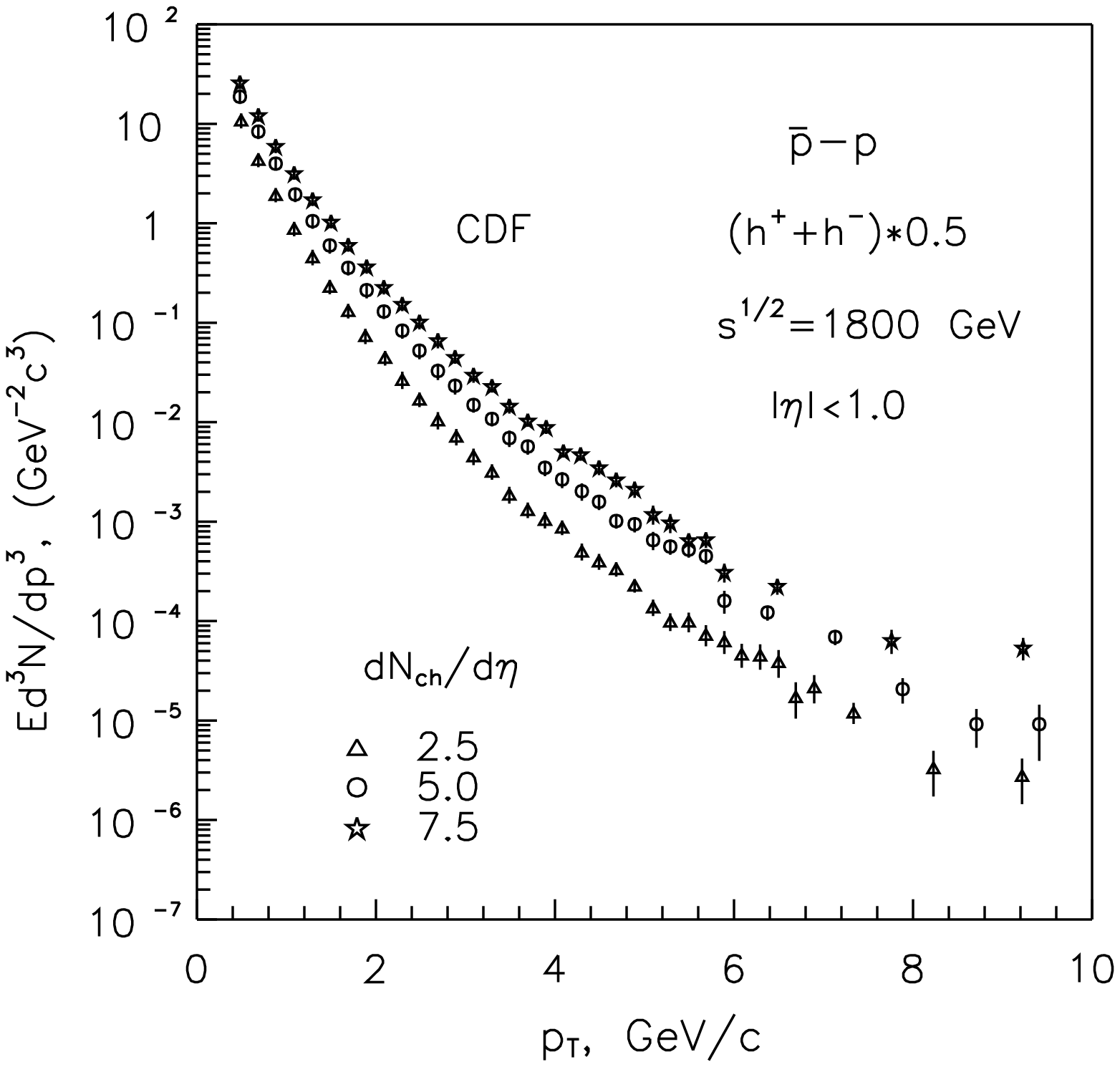}{}}
\vskip -1.cm
\hspace*{0.cm} a) \hspace*{8.cm} b)\\[0.5cm]
\end{center}

\vskip 5cm

\begin{center}
\hspace*{-2.5cm}
\parbox{5cm}{\epsfxsize=5.cm\epsfysize=5.cm\epsfbox[95 95 400 400]
{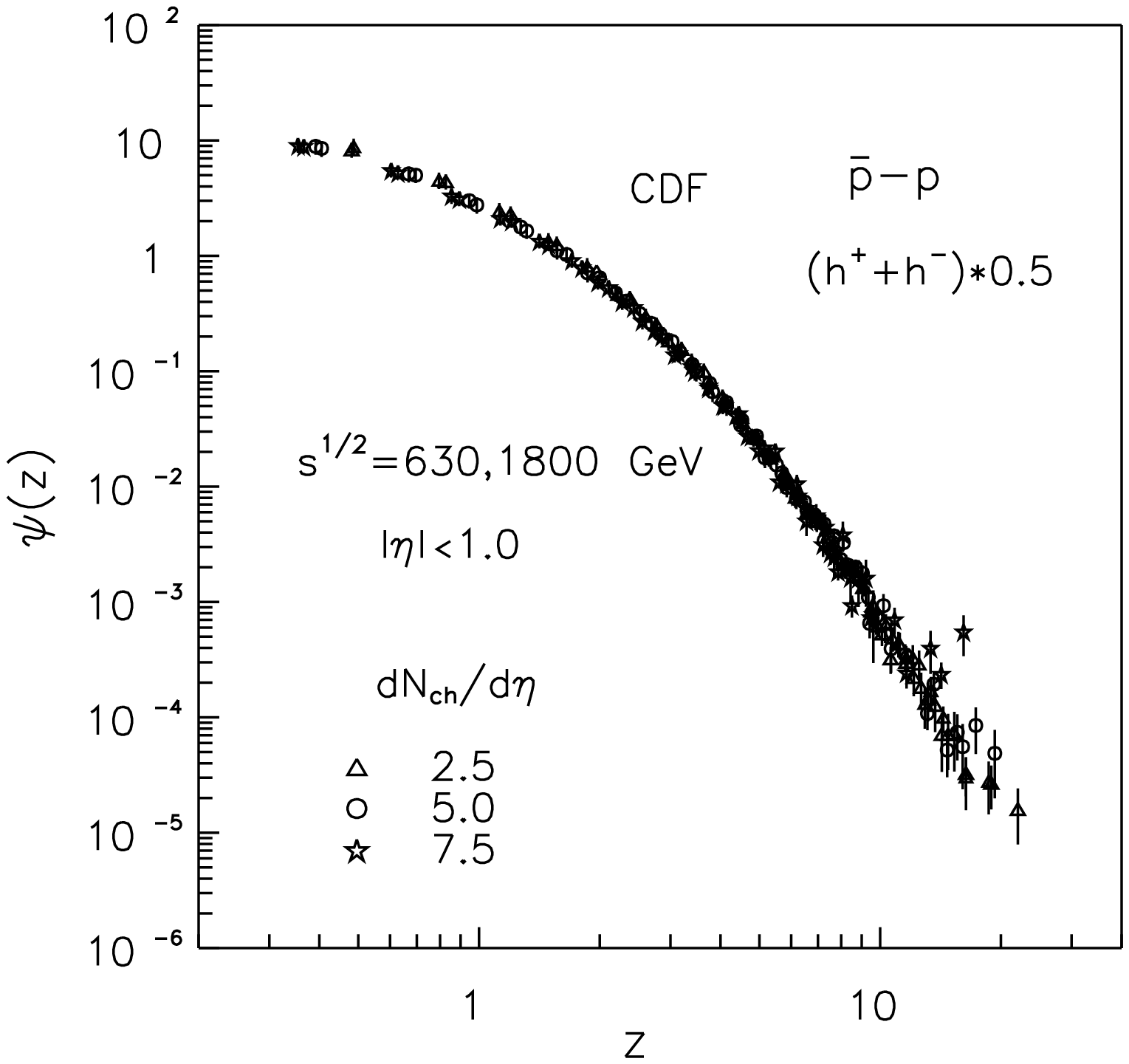}{}}
%\hspace*{3cm}
%\parbox{5cm}{\epsfxsize=5.cm\epsfysize=5.cm\epsfbox[95 95 400 400]
%{hchmul0.eps}{}}
\vskip -1.cm
\hspace*{0.cm} c) \\[0.5cm]
\end{center}

{\bf Figure 3.}
 Multiplicity dependence of charged hadron spectra
  in $\bar pp$ collisions  at $\sqrt s=630$ (a) and 1800~GeV (b).
 Experimental data are obtained by the CDF Collaboration \cite{CDF}.
(c) The corresponding scaling function $\psi(z)$.

%\end{document}

% *************    4(a,b) 5(a,b)  *************************
\newpage
\begin{minipage}{4cm}

\end{minipage}

\vskip 4cm
\begin{center}
\hspace*{-2.5cm}
\parbox{5cm}{\epsfxsize=5.cm\epsfysize=5.cm\epsfbox[95 95 400 400]
{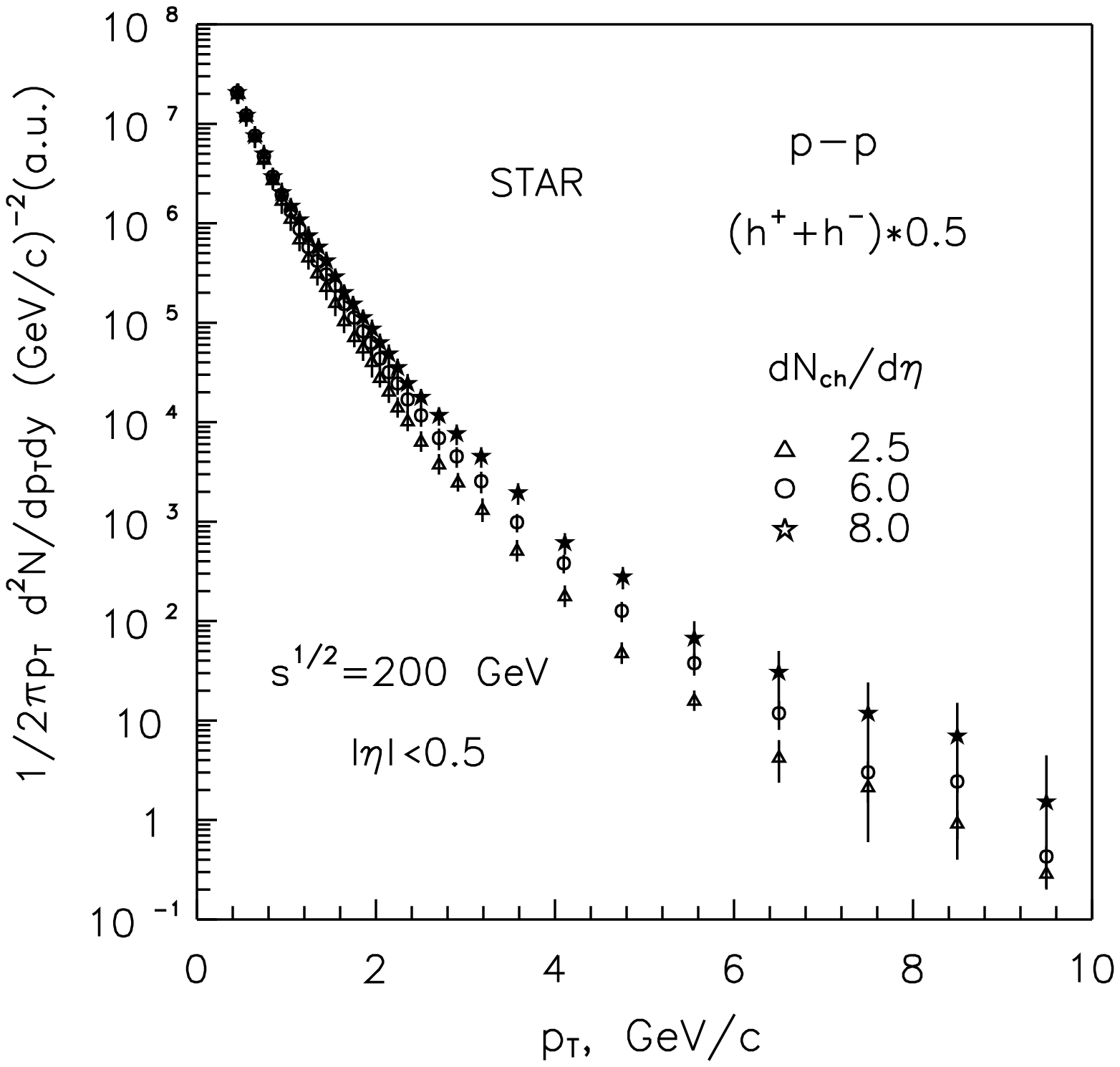}{}}
\hspace*{3.cm}
\parbox{5.cm}{\epsfxsize=5.cm\epsfysize=5.cm\epsfbox[95 95 400 400]
{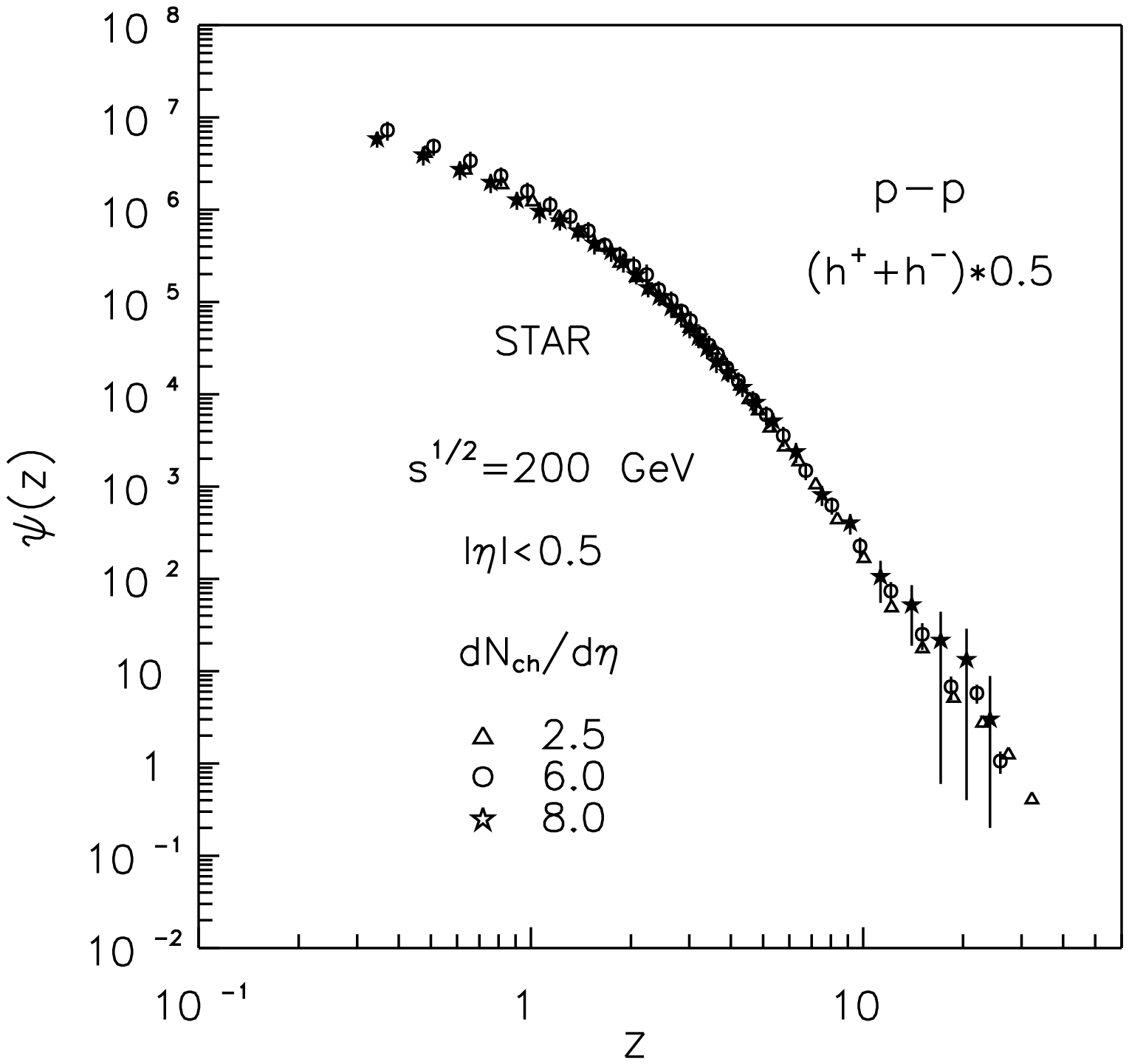}{}}
\vskip -1cm
\hspace*{0.1cm} a) \hspace*{8.cm} b)\\[0.5cm]
\end{center}

{\bf Figure 4.}
(a) Multiplicity dependence of charged hadron spectra
  in $ pp$ collisions  at $\sqrt s=200$~GeV.
 Experimental data are obtained by the STAR Collaboration \cite{STAR}.
(b) The corresponding scaling function $\psi(z)$.

\vskip 5cm

\begin{center}
\hspace*{-2.cm}
\parbox{6cm}{\epsfxsize=6.cm\epsfysize=6.cm\epsfbox[95 95 400 400]
{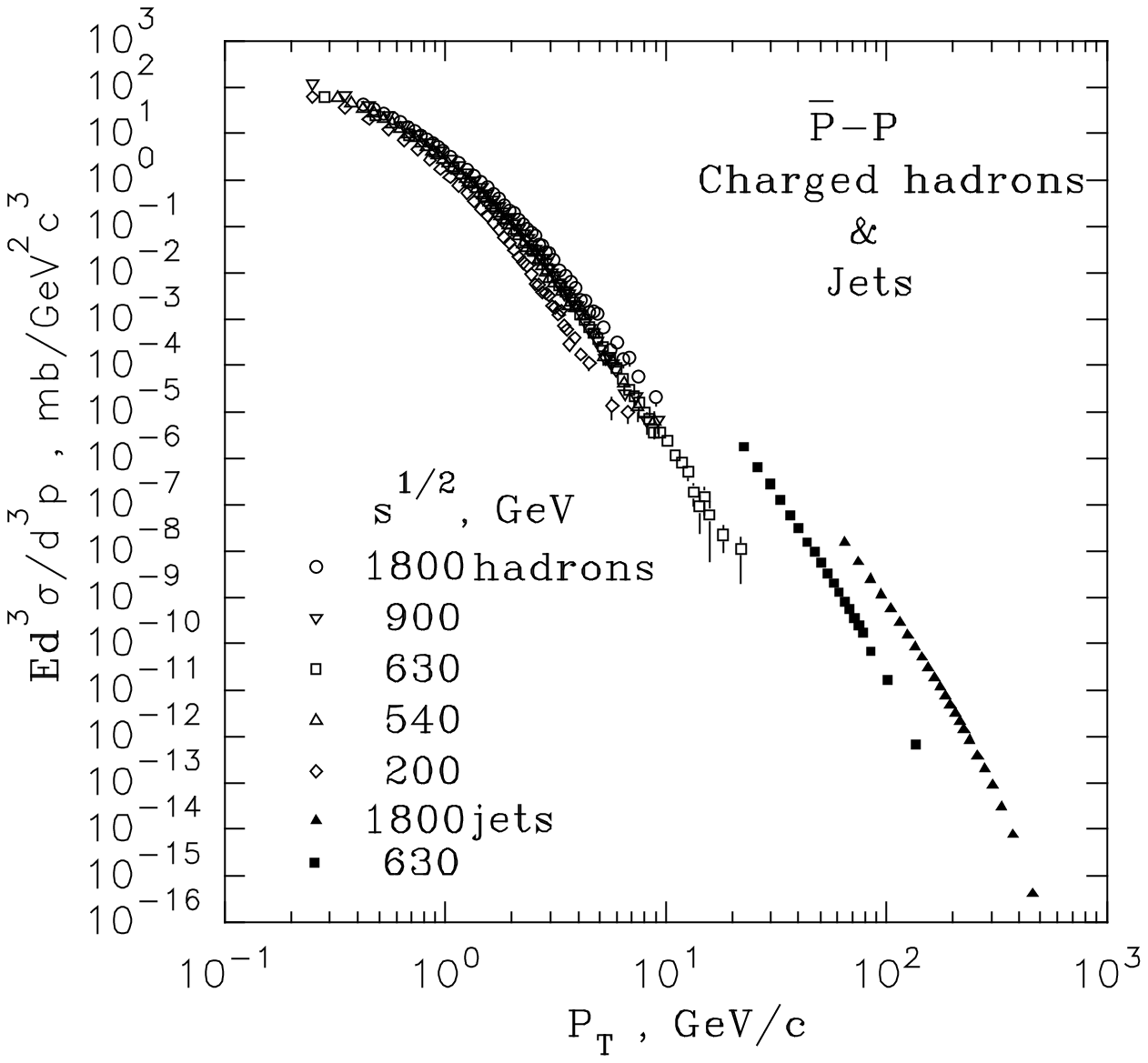}{}}
\hspace*{3.cm}
\parbox{6cm}{\epsfxsize=6.cm\epsfysize=6.cm\epsfbox[95 95 400 400]
{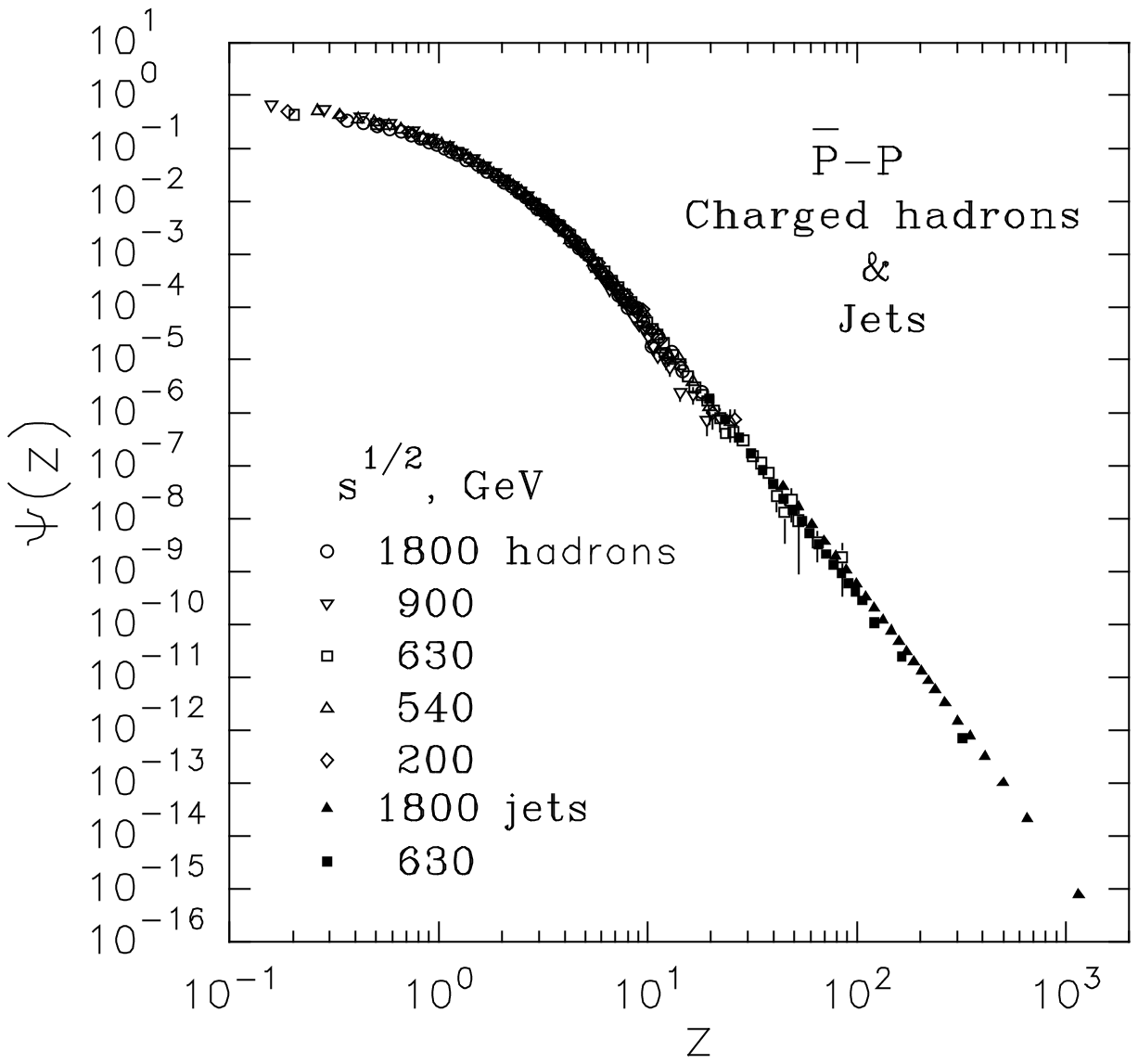}{}}
\vskip -2.cm
\hspace*{0.cm} a) \hspace*{8.cm} b)\\[0.5cm]
\end{center}

{\bf Figure 5.}
(a) Charged hadron and  jet spectra
  in minimum bias $\bar pp$ collisions
  as a function of collision energy.
 Experimental data are taken from \cite{hpm} and \cite{D0jet}.
(b) The corresponding scaling function $\psi(z)$.


\vskip 0.5cm
\begin{thebibliography}{99}


\bibitem{Quarkcom}
E. Eichten, K. Lane, M. Peskin, Phys. Rev. Lett. {\bf 50}, 811 (1983).\\
E. Eichten, I. Hinchliffe, K. Lane, C. Quigg, Rev. Mod. Phys. {\bf 56}, 4 (1984).

\bibitem{Extradim}
I. Antoniadis, In: Proceedings of European School of High-Energy Physics,
Beatenberg, Switzerland, 26 August - 8 September, 2001 (Editors: N.Ellis,
J.March-Russul) p.301

\bibitem{Blackhole}
C.G. Lester, In: Proceedings of Advanced Studies Institute on "Physics at LHC",
Czech Republic, Prague, July 6-12, 2003 (Editors: M Finger,A.Janata, M.Virius)
A303.

\bibitem{Fracspace}
%\bibitem{Nottale}
L. Nottale, {\it Fractal Space-Time and Microphysics} (World Sci., Singapore, 1993).\\
%\bibitem{Mandelbrot}
B. Mandelbrot, {\it The Fractal Geometry of Nature} (Freeman, San Francisco, 1982).
%\bibitem{Imr}
%I. Zborovsk\'{y}, hep-ph/0311306.

\bibitem{Feynman}
R.P. Feynman, Phys. Rev. Lett. {\bf 23}, 1415 (1969).

\bibitem{Bjorken}
J.D. Bjorken, Phys. Rev. {\bf 179}, 1547 (1969);
J.D. Bjorken, and E.A. Paschanos, Phys. Rev.
{\bf 185}, 1975 (1969).

\bibitem{Bosted}
P. Bosted {\it et al.}, Phys. Rev. Lett. {\bf 49}, 1380 (1972).

\bibitem{Benecke}
J. Benecke  {\it et al.},  Phys. Rev.  {\bf 188}, 2159 (1969).

\bibitem{Baldin}
A.M. Baldin,
Sov. J. Part. Nucl. {\bf 8}, 429 (1977).

\bibitem{Stavinsky}
 V.S. Stavinsky,
Sov. J. Part. Nucl. {\bf 10}, 949 (1979).


\bibitem{Leksin}
G.A. Leksin: Report No. ITEF-147, 1976; G.A. Leksin: in
{\it Proceedings of the XVIII International Conference on High
Energy Physics}, Tbilisi, Georgia, 1976, edited by N.N.
Bogolubov {\it et al.}, (JINR Report No. D1,2-10400, Tbilisi,
1977),p. A6-3.

\bibitem{KNO}
Z. Koba, H.B. Nielsen, and P. Olesen,
Nucl. Phys.  {\bf B40}, 317 (1972).

\bibitem{Matveev}
  V.A. Matveev, R.M. Muradyan, and A.N. Tavkhelidze,
  Part. Nuclei {\bf 2}, 7 (1971);
  Lett. Nuovo Cim. {\bf 5},  907 (1972);
  Lett. Nuovo Cim. {\bf 7}, 719 (1973).

\bibitem{Brodsky}
  S. Brodsky, and G. Farrar,
  Phys. Rev. Lett. {\bf 31}, 1153 (1973);
  Phys. Rev.  {\bf D11}, 1309 (1975).


\bibitem{Z}
I.Zborovsk\'{y}, Yu.A.Panebratsev, M.V.Tokarev, and G.P.\v{S}koro,
Phys. Rev. {\bf D 54}, 5548 (1996);
I.Zborovsk\'{y}, M.V.Tokarev, Yu.A.Panebratsev, and G.P.\v{S}koro,
Phys. Rev. {\bf C59}, 2227 (1999);
M.V.Tokarev, T.G.Dedovich, Int. J. Mod. Phys. {\bf A15}, 3495 (2000);
M.V.Tokarev, O.V.Rogachevski, T.G.Dedovich,
J. Phys. G: Nucl. Part. Phys. {\bf 26}, 1671 (2000);
M.V.Tokarev, O.V.Rogachevski, and T.G.Dedovich,
Preprint No. E2-2000-90, JINR (Dubna, 2000);
M.Tokarev, I.Zborovsk\'{y}, Yu.Panebratsev, and G.Skoro,
Int. J. Mod. Phys. {\bf A16}, 1281 (2001);
M.Tokarev, hep-ph/0111202;
M.Tokarev, D.Toivonen, hep-ph/0209069;
G.P.Skoro, M.V.Tokarev, Yu.A.Panebratsev, and I.Zborovsk\'{y}, hep-ph/0209071;
M.V.Tokarev, G.L.Efimov, and D.E.Toivonen, Physics of Atomic Nuclei, {\bf 67}, 564 (2004).
M.Tokarev, Acta Physica Slovaca, {\bf 54}, 321 (2004); M.V.Tokarev, and T.G.Dedovich,
Physics of Atomic Nuclei, {\bf 68}, 404 (2005).


\bibitem{UA1}
 G. Arnison {\it  et al.}, Phys. Lett.  {\bf B118}, 167 (1982).


\bibitem{E735}
  T. Alexopoulos  {\it et al.}, Phys. Lett. {\bf B336}, 599 (1994).


\bibitem{CDF}
   D. Acosta {\it et al.}, Phys. Rev. {\bf D65},  072005 (2002).


\bibitem{STAR}
   J.E. Gans,  PhD Thesis,
   Yale University, USA (2004).

\bibitem{hpm}
   G. Arnison {\it et al.}, Phys. Lett. {\bf  B118}, 167 (1982).\\
   F. Abe {\it et al.}, Phys. Rev. Lett. {\bf 61},  1819 (1988).\\
   C. Albajar {\it et al.}, Nucl. Phys. {\bf B335},  261 (1990).\\
   G. Bocquet {\it et al.}, Phys. Lett. {\bf B366}, 434 (1996).
\bibitem{D0jet}
   B. Abbott {\it et al.}, Phys. Rev. {\bf D64},  032001 (2001).



\end{thebibliography}
\end{document}